# A Bose-Einstein Model of Particle Multiplicity Distributions

**A.Z.Mekjian [1,2,3], T. Csörgő [3] and S.Hegyi [3]**

[1] Department of Physics and Astronomy, Rutgers University, Piscataway, N.J. 08854
[2] California Institute of Technology, Kellogg Radiation Lab., Pasadena, CA 91106
[3] MTA KFKI RMKI, H-1525 Budapest 114 PO Box 49, Hungary

### Abstract

A model of particle production is developed based on a parallel with a theory of Bose-Einstein condensation and similarities with other critical phenomena such as critical opalescence. The role of a power law critical exponent $\tau$ and Levy index $\alpha$ are studied. Various features of this model are developed and compared with other commonly used models of particle production which are shown to differ by having different values for $\tau, \alpha$. While void scaling is a feature of this model, hierarchical structure is not a general property of it. The value of the exponent $\tau = 2$ is a transition point associated with void and hierarchical scaling features. An exponent $\gamma$ is introduced to describe enhanced fluctuations near a critical point. Experimentally determined properties of the void scaling function can be used to determine $\tau$.

PACS numbers: 25.75Dw, 25.75Gz, 25.75Nq

## 1.Introduction

In this paper we study particle multiplicity distributions based on a model that parallels a theory of Bose Einstein condensation and has a connection to other critical phenomena such as critical opalescence. Their presently exists considerable interest in Bose-Einstein condensation for several reasons. One of the main reasons for the recent interest is the actual observation of the Bose-Einstein condensation phenomena in laser traps [1-2]. Bose-Einstein correlations are fundamental in the Hanbury-Brown Twiss (HBT) effect [3]. HBT correlations have been used to measure source sizes in particle and heavy ion collisions [4,5]. Parallels of Bose-Einstein condensation have been developed in the theory of phase transitions in medium energy heavy ion collisions [6]. Here the condensation of the macroscopic liquid state from nucleation processes in a liquid/gas phase transition is analogous to the macroscopic formation of the ground state in the condensation process. Application of Bose-Einstein condensation phenomena have also appeared in the theory of linkage and the internet [7,8]. Results based on a pion laser model [9,10] have similarities with condensation in harmonic oscillator traps. In a pion laser model and other models of particle multiplicity distributions, interest focuses on count probability distributions and fluctuations. Bose-Einstein statistics enhance emission probabilities and lead to statistical correlations in phase space. Several models of photon count distributions from lasers can be found in ref [11]. Phenomenological models of hadron multiplicities, that are based on photon count distributions, have been applied to understanding high energy collisions[12]. These models have also been extended [13,14]. Comparisons will be made with these other models and the one developed in this paper. A main element, associated with a power law critical exponent $\tau$, distinguishes various models in our generalized approach which leads to a unified description of count distributions. We also explore a connection with the phenomena of critical opalescence. Critical opalescence occurs at the end point of a first order phase transition where a strong enhancement in density fluctuations is seen. At this endpoint, the phase transition becomes second order. Similarly. in a heavy ion collision a strong enhancement in particle fluctuations may occur associated with a possible critical point. An exponent $\gamma$ will be introduced to describe this enhancement in particle fluctuations near a critical point.

In particle multiplicity distributions features related to scaling properties play an important role. Examples of such scaling features are KNO scaling[15], void scaling[16], and hierarchical structure relations[17]. Some of these distributions have a connection with Levy distributions in probability theory [18] which also has important applications in many areas of physics and the natural sciences. A Levy distribution is non-Gaussian with an asymptotic power law behavior. The exponent associated with the asymptotic fall off of the distribution is the Levy index. Infinite moments of the probability distribution occur for pure power law decays in the probability distribution. Stable Levy distributions have finite moments where the stabilization is established either by having an exponential factor besides the power law part or by truncation of the distribution itself to a finite number of terms. One important feature of pure power laws is the lack a length scale. The relation of this behavior to physics occurs at critical points in the thermodynamic behavior of systems. At a critical point length scales disappear and thermodynamic quantities such as specific heat and compressibility become singular. The two point correlation function also diverges. A second order phase transition has associated with it a set of critical exponents that characterize these various properties. They include: a) the exponent associated with the power law fall off of droplet sizes in a liquid/gas phase transition or of clusters in percolation theory at the onset of the formation of the infinite cluster; b) exponents associated with the divergences of various thermodynamic quantities such as the compressibility, specific heat; c) an exponent associated with a correlation length and the scale invariance of the system. An example pertinent to this paper is the $\lambda$ transition in liquid helium. The name $\lambda$ transition comes from the singular shape of the specific heat which has a sharp rise as the critical temperature is approached from below and a sharp fall off above the critical temperature with increasing $T$. Bose-Einstein condensation plays a key role in this phenomena [19,20]. The simplest model of non-interacting bosons [19,20] confined to a volume $V$ gives a cusp for the behavior of the specific heat at constant volume. Levy distributions have also been useful in understanding Bose-Einstein correlations [21]. For a discrete probability distribution where the probability for $N$, or $P_N$, falls as a power as $\sim 1/N^{\tau}$, the first moment which gives the $< N >$ is finite if the index or exponent $\tau > 2$. If $1 < \tau \leq 2$, then the first moment also diverges. For this second case with $1 < \tau \le 2$, $\tau$ is related to the Levy index $\alpha$ of a continuous distribution as will be discussed. However, in most statistical physics problems and in percolation theory the power law fall of some quantity has $\tau > 2$. The review of ref[25] also has many examples of places outside of physics where the index $\tau$ is in the range $2 < \tau \le 3$. These facts lead us to explore a model with a continuous range of values of $\tau$. We also show how some popular models of particle production differ only by the value of the exponent $\tau$. The value $\tau = 2$ is found to be a transition point in the theory. The void scaling properties and hierarchical structure features differ for $\tau < 2$ and $\tau > 2$. The Bose-Einstein model has $\tau = 5/2$, while the critical opalescence model has $\tau = 2 + 1/3$. A frequently used count probability distribution is the negative binomial distribution which has $\tau = 1$ and it is exponentially stabilized. The negative binomial distribution has non-Poissonian fluctuations.

Recently [26], generating functions have also been used in the discussion of particle multiplicity distributions in field theories coupled to strong time dependent external sources. An important example of such a field theory is the Color Glass Condensate as discussed in ref [26]. The distribution obtained is shown to be non-Poissonian even in the limit where classical approximations to the equations of motion are valid. The particle multiplicity distribution of several phenomenological models of particle production can also be found in the reviews of ref[27,28]. The physics report of Dremin and Gary illustrates how phenomenological approaches to multiplicity distributions based on the use of generating functions and probability theory can be useful in understanding particle emission. A particular emphasis in these physics reports is the scaling behavior of the factorial moments and non-Poissonian fluctuations and the behavior of the cumulant moments with increasing order. Dremin, in ref[29], has proposed a simple analytic model of particle production based on $\varphi^3$ theory with strong external sources. Properties of his analytic model resemble some features associated with a phase transition.

We should mention the usefulness of void scaling relations and hierarchical structure in the study of the large scale structure of the universe. A recent study of such observed structure in terms of void scaling and hierarchical structure can be found in ref[22]. Some early work on these features as applied to astrophysics and cosmology are in ref[16,17,23,24]. Hierarchical models play a major role in the description of galaxy clusters. It is therefore of general interest to establish the conditions for having hierarchical scaling and how this scaling influences properties of void distributions.

The determination of critical exponents is important both theoretical and experimentally. Properties of the void scaling function could be used to determine the exponent $\tau$. As discussed in this manuscript, the shape of the void scaling function with increasing scaling parameter and its end point value can be used for this purpose.

## 2. General Results

In this section we summarize some results that will be used. Appendix A also contains some of the connections which are motivated by considerations of Bose-Einstein condensation. The count probability distribution for n-particles [27,28] can be obtained from a generating function $G(u)$

$$G(u) = \sum_{N=1}^{\infty} P_N u^N$$

(1)

The $G(u)$ is related to combinants $C_k$ through a connection that reads

$$G(u) = Exp(\sum_{k=1}^{\infty} C_k (u^k - 1))$$

(2)

The unnormalized factorial moments $F_q$ of $P_N$ are obtained from $G(u)$ using

$$F_q = \left(d^q G(u) / du^q\right)_{z=1} = \sum_{N=0}^{\infty} N(N-1)...(N-q+1)P_N$$

(3)

The unnormalized cumulants $K_q$ are obtained from $Ln(G(u))$ through

$$K_q = \left(d^q (Ln(G(u)) / du^q\right)_{z=1}$$

(4)

The cumulants are connected to factorial moments of the combinants through the connection

$$K_q = \sum_{k=1}^{\infty} k(k-1)...(k-q+1)C_k$$

(5)

In turn, the factorial moments $F_q$ and cumulants $K_q$ obey a recurrence relation that is

$$F_q = \sum_{n=0}^{q-1} \frac{(q-1)!}{n!(q-n-1)!} K_{q-n} F_n$$

(6)

The mean multiplicity $< N >$ is related to various quantities just mentioned as

$$< N > = F_1 = K_1 = \sum_{k=1}^{\infty} kC_k$$

(7)

The fluctuation $<N^2>-<N>^2=<(N-<N>)^2>=\Sigma(N-<N>)^2 P_N$ is the second moment of the $C_k$ distribution:

$$<N^2>-<N>^2=\sum_{k=1}^{\infty}k^2 C_k$$

(8)

while the skewness $<(N-<N>)^3>$ is the third moment:

$$<(N-<N>)^3>=\sum_{k=1}^{\infty}k^3 C_k$$

(9)

The $F_2=<N(N-1)>$ while $K_2=<N^2>-<N>^2-<N>=F_2-<N>^2=F_2-(F_1)^2$. We will write the fluctuation as

$$<N^2>-<N>^2=<N>+B<N>^2$$

(10)

so that $K_2=B<N>^2$. The void probability is $P_0=1/Exp(\Sigma C_k)$ and involves the zeroth moment of the $C_k$ distribution. Associated with $P_0$ is the void function $\chi$ defined as

$$\chi\equiv -Ln(P_0)/<N>=\sum_{k=1}^{\infty}C_k/\sum_{k=1}^{\infty}kC_k$$

(11)

Of interest is a possible scaling behavior of $\chi$ with the variable $B<N>=K_2/<N>$.

Another scaling behavior of importance is a property associated with normalized cumulants $k_q=K_q/<N>^q$. An hierarchical scaling relation is present when a simple power structure is present that reads

$$k_q=A_q(k_2)^{q-1}$$

(12)

The $A_q$ depends only on the order $q$ in this connection of $k_q$ to the $q-1$ power of $k_2$.

To further establish some of the results just noted two often quoted examples will now be given before we proceed with our generalized model of the next section. These examples are the Poisson distribution

$$P_N = \frac{<N>^N}{N!} Exp(-<N>)$$
(13)

and the negative binomial distribution

$$P_N = \frac{(N+x-1)!}{N!(x-1)!}\left(\frac{<N>/x}{1+<N>/x}\right)^N \left(\frac{1}{1+<N>/x}\right)^x = \frac{(N+x-1)!}{N!(x-1)!} p^x (1-p)^N$$
(14)

with $p = 1/(1+<N>/x)$.

The combinants $C_k$ for a Poisson are $C_1 =<N>$, $C_{k\neq 1} = 0$. For a Poisson $<N^2> - <N>^2 =<N> = K_1$. The $K_2 = 0$ and also all higher $K_q = 0$. The $F_q =<N>^q$. For a Poisson distribution, the $\chi = 1$ and is independent of the variable $B<N>$ or $K_2/<N>$.

The combinants for a negative binomial are a distribution given by $C_k = xz^k/k$. The $<N> = xz/(1-z)$ and $<N^2> - <N>^2 =<N> + (<N>^2/x) =<N>(1+<N>/x)$. If $z \to 1$, $<N> \to \infty$. A value of $z$<1 exponentially truncates the distribution of $C_k$ and leads to finite moments of it. The void scaling relation is $\chi = Ln(1+B<N>)/B<N>$ and an hierarchical structure relation $k_q = A_q (k_2)^{q-1}$ exists with $A_q = (q-1)!$. The $K_2 =<N>^2/x$ and $k_2 = K_2/<N>^2 = 1/x$. The factorial moments are given by $F_q = (<N>/x)^q x(x+1)...(x+q-1)$ and the normalized factorial moments are $f_q \equiv F_q/<N>^q = x(x+1)...(x+q-1)/x^q$ or $f_q = \Gamma(x+q)/(\Gamma(x)x^q)$.

The results of appendix A show a parallel between the generating function/particle multiplicity distribution connection and the grand canonical ensemble/canonical ensemble partition function relationship. Namely, the grand canonical ensemble $Z_{gc}(z)$ and canonical ensemble $Z_A$, for a system of $A$ particles with $A = 0,1,2,....\infty$, are connected by the relation

$$Z_{gc} = \sum_{A=0}^{\infty} Z_A z^A \qquad (15)$$

where $z = \exp(\beta\mu)$ is the fugacity, $\beta = 1/T$ is the inverse temperature, and $\mu$ is the chemical potential. Studies of Bose-Einstein condensation outlined in this appendix motivated a choice for the combinant $C_k$ given in the next section. A summary of the main ideas of appendix A is given in the section A.4.

## 3. Hypergeometric model approximation and BE condensation

3.1 General case $F(a,b;c,z)$

We now introduce a model for particle multiplicity distributions that is an approximate representation of BE condensation. A parallel with critical phenomena and critical opalescence will be developed. The model is analytic and also contains all the features of BE correlations and condensation phenomena. The model is based on an extension of a hypergeometric model ${}_2F_1(a,1;2,z)$ used in ref[14] to study particle multiplicity distributions. The model involves the hypergeometric function ${}_2F_1(a,b;c,z)$ which appears in evaluating the $Z_{gc}$ or generating function $G(u)$ for a specific form of $C_k$. The subscripts 2,1 in ${}_2F_1$ will be omitted to shorten the notation. A hypergeometric model will have

$$Z_{gc} = \exp(xzF(a,b;c,z))$$

(16)

with

$$F(a,b;c,z) = \sum_{k=0}^{\infty} \frac{[a]_k [b]_k}{[c]_k k!} z^k$$

(17)

The $[a]_k = a(a+1)(a+2)....(a+k-1)$. The $k = 0$ term in $F(a,b;c,z)$ is 1. The $C_k$ is then given by:

$$C_k = x \frac{[a]_{k-1} [b]_{k-1}}{[c]_{k-1} (k-1)!} z^k$$

(18)

The $x$ and $z$ are treated as variables in this approach. Section 4 contains some examples where $x$ and $z$ are specified in particular models. The $x$ and $z$ determine the mean multiplicity and its fluctuation. This feature was already illustrated for the negative binomial where $< N >= xz/(1-z)$ and the $x$ is the negative binomial parameter that appears in $P_N$ of eq(14) and in the fluctuation $< N^2 > - < N >^2 =< N > (1+ < N > / x)$. The results of eq (13-15) will be the basis of our BE model of count distributions with a specific choice for $a,b,c$ which is $a = 1/2,\ b = 1, c = 3$. However, the more general case $a, b = 1, c = 3$ will also be discussed to some extent and another particular choice $a = 2/3,\ b = 1, c = 3$ leads to results that have some connection with the phenomena of critical opalescence. Using these equations we are able to study this model in an analytic way where the analysis is simple. This feature also will allow us to study some specific quantities such as void scaling and hierarchical structure features that may be present in this model. By varying $a,b,c$ various critical exponents $\tau$ can be generated.

As mentioned, in ref[14] results for the case of general $a$ and $b = 1, c = 2$ or $F(a,1;2,z)$

were presented. For $a$=1 the associated particle multiplicity distribution of $F(a,1;2,z)$ was the negative binomial (NB) distribution. For $a$=1/2, the associated multiplicity distribution of $F(a,1;2,z)$ is another model taken from quantum optics[30,11] due to Glauber and applied to particle multiplicity distributions in ref[31]. For $a$=1, the $C_k = xz^k / k$. For $a$=1/2, the $C_k = (1/k)(2(k-1))!/((k-1)!(k-1)!2^{2(k-1)})xz^k$ which, for large $k$, falls as $C_k \sim xz^k / k^{3/2}$. The exponent $\tau$ is the power in the $1/k^\tau$ behavior of $C_k$ for large $k$. Then $\tau = 1$ for $a$=1 and $\tau$ =3/2 for $a = 1/2$ for the case $F(a,1;2,z)$. For general $a$ and with $b,c$ still restricted by $b = 1, c = 2$, the $C_k \sim xz^k / k^{2-a}$ for large $k$ so that $\tau = 2 - a$. The $a \to 0$ limit has $C_k$=0 for $k$=2,3,.., except $C_1 = xz$. The resulting distribution is a Poisson with $< N >= xz$. The case $a = 1/2, b = 1, c = 2$ has connections with a Feynman/Wilson gas [14] . We now turn our attention to the particular case $a = 1/2, b = 1, c = 3$ where exact analytic results can be easily obtained.

3.2 Specific case $a = 1/2, b = 1, c = 3$; the Bose-Einstein parallel, $\tau = 5/2, \gamma = 1/2$ .

Since many distributions have $\tau$ in the range $2 < \tau < 3$, we have extended the hypergeometric model to ${}_2F_1(a,1;3,z)$, where b=1 and c=3. Moreover, we will also specialize the discussion to $a = 1/2$ where $C_k \sim xz^k / k^{5/2}$ for large $k$ and $\tau$ =5/2. The more general case of a continuous $a$ in the range $0 < a < 1$, but with $b = 1, c = 3$ has a power law exponent $\tau = 3 - a$. A brief discussion of this general case will also be given.

In BE condensation the expansions in fugacity z involves the functions

$$G_r(z) = \sum_{k=1}^{\infty} z^k / k^r$$

(19)

In d=3 dimensions, the exponent r=1+d/2=5/2 appears in the grand canonical partition function and r=3/2 appears in the expression for $< N >$ as given in eq(A.2) and eq(A.3a). At condensation: $z$ =1 and $G_{5/2}(z=1) = \varsigma(5/2) = 2.61$ and $G_{3/2}(z=1) = \varsigma(3/2) = 1.34$. The grand canonical partition is $\exp(xG_{5/2}(z))$ where $x = V / \lambda^3 = V(2\pi mT)^{3/2} / h^3$, as developed in Appendix A. The $< N >= xG_{3/2}(z)$ was used to determine the critical $T = T_C$ or $V = V_C$ by allowing $x$ to vary, either at fixed $T$ or $V$, until the $z = 1$ condensation point was reached. By comparison, the hypergeometric BE model $F(1/2,1;3,z)$ has a simpler closed form expression that reads

$$Z_{gc} = \exp(xzF(1/2,1;3,z)) = \exp((4/3)x\frac{(1-z)^{1/2} - 2(1-z) + 1}{1 + (1-z)^{1/2}})$$
$$= \exp((4/3)x\frac{(1-t)(2t+1)}{1+t})$$

(20)

where $t=(1-z)^{1/2}$. At the point $z=1, t=0$, the $Z_{gc}=\exp((4/3)x)$, which can be compared to the BE result which has the 4/3=1.33 factor replaced with $\varsigma(5/2)=1.34$. In the hypergeometric model the $<N>$ can be obtained from $<N>=z\partial xzF(1/2,1:3,z)/\partial z$ and is

$$<N>=(4/3)x\frac{(1-(1-z)^{1/2})(2+(1-z)^{1/2})}{(1+(1-z)^{1/2})}=(4/3)x\frac{(1-t)(2+t)}{(1+t)}$$

(21)

The condensation condition now reads $<N>=(8/3)x_C$ so that the $\varsigma(3/2)=2.61$ is replaced with 8/3=2.67. The behavior of the fugacity $z$ and associated chemical potential $\mu$ is obtained numerically in the BE condensation picture by solving eq(1) for $z$ as a function of $<N>/x$. In the approximate model of this section a simple quadratic equation for $t$ has to be solved using eq(21). Defining $\xi=(3/4)<N>/x$, the solution to eq(21) gives

$$t=(1-z)^{1/2}=-(1/2)(1+\xi)+(1/2)((1+\xi)^2-4(\xi-2))^{1/2}$$

(22)

At condensation, $\xi=2$, and $t=0, z=1$. The variance $<N^2>-<N>^2$ and a related quantity $<N^2>-<N>^2-<N>\equiv B<N>^2$ can also be easily calculated. Of special interest is the associated function $B<N>=(<N^2>-<N>^2-<N>)/<N>$ which is the ratio of the second factorial moment to the first factorial moment, with the later being $<N>$. The $B<N>^2$ term gives the departures from Poisson statistics. The $B<N>$ is

$$B<N>=\frac{(1-t)(3+t)}{t(2+t)}=\frac{(1-(1-z)^{1/2})(3+(1-z)^{1/2}}{(1-z)^{1/2}(2+(1-z)^{1/2})}$$

(23)

For $z\approx 1$, the $B<N>$ will diverge as $1/\sqrt{1-z}$. An exponent $\gamma$ can be introduced to describe this divergence or that of $\delta N^2$ which also goes as $1/\sqrt{1-z}$ in this case. Since $\delta N^2$ is related to $\kappa_T$, we will use $\delta N^2$ to define $\gamma$ through the relation $\delta N^2\sim 1/(1-z)^{\gamma}$. The value $\gamma=1/2$ applies for the Bose-Einstein model.

Another function of interest, which comes from cosmology and particle multiplicity distributions [16,17,22-24], is a void function $\chi$ defined by eq(11) and rewritten here in terms of clan variables as $\chi=N_C/<N>=1/n_C$. The $N_C=\sum C_k$ is referred to as the number of clans or number or Poisson clusters, while $n_C\equiv<N>/N_C$ is the average number of particles per clan or Poisson cluster. At the condensation point: $t=0, N_C=4/3, <N>=8/3$. The void function $\chi$ scales with $B<N>$ as

$$\chi = 2 - \frac{3}{1 + (1 + \frac{3}{1 + 2 < N > B})^{1/2}}$$

(24)

This feature is called void scaling. The limiting value of $\chi$ is ½ as $B < N > \to \infty$ for this particular hypergeometric model , which is also the value of $N_C / < N >$. The value of $< N > / N_C$ in terms of $t$ or $z$ is

$$\frac{< N >}{N_C} = n_C = \frac{2 + t}{1 + 2t} = \frac{2 + (1 - z)^{1/2}}{1 + 2(1 - z)^{1/2}}$$

(25)

For $C_k$ which fall slower than or equal to $1/k^2$, $< N > / N_C \to \infty$ and $\chi \to 0$ as $B < N > \to \infty$ or $z \to 1$. The $\chi \to 0$ is a characteristic behavior of any $F(a,1;,2,z)$ model. Thus the exponent $\tau = 2$ that characterizes this power law behavior is a transition point , not only in whether BE condensation happens at zero or non-zero values of $T$ as discussed in the appendix, but also in the asymptotic behaviors of $\chi$ and $< N > / N_C = n_C$ or $N_C / < N >$. For $\tau > 2$, the $N_C$ and $< N >$ are both finite. The value of $B < N > \to \infty$ for $\tau \leq 3$ with $z \to 1$ and $\chi \to N_C / < N >$, a non-zero number. For $1 < \tau \leq 2$, $N_C$ is finite, $< N > \to \infty, B < N > \to \infty$, $N_C / < N > \to 0$ and $\chi \to 0$ as $B < N > \to \infty$.

As noted in the introduction, the determination of critical exponents is a major endeavor both theoretically and experimentally. The results just presented regarding the behavior of the void scaling function $\chi$ with $B < N >$ also allow for an experimental determination of it. The simplest observation is whether $\chi \to 0$ or a non-zero value $\chi \neq 0$ with increasing $B < N >$ or cumulant variable $K_2 / < N >$. For $\tau \leq 2$, $\chi \to 0$, and for $\tau > 2$, $\chi$ does not $\to 0$. For $\tau > 2$, the endpoint value of $\chi = N_C / < N > = \Sigma C_k / \Sigma k C_k$ is useful for determining $\tau$ since $\chi \to 1 - a = \tau - 2$. These results apply for the hypergeometric models based on $xzF(a,1;3,z)$.

An interesting relationship may also exist for the factor moments of the distribution of the $C_k$. In the literature the relationship connects higher factorial moments to the second factorial moment and is known as hierarchical structure and it will be made more precise below. The q'th factorial moment of the $C_k$ 's is $K_q = \sum k(k-1)(k-2)...(k-q+1)C_k$. The $K_q$ are called cumulants. The cumulants $K_q$ can be obtained from $K_q = xz^q (\partial^q / (\partial z)^q (zF(1/2,1;3,z))$ which leads to the result:

$$K_q = xz^q \{q(\frac{\Gamma(1/2 + q - 1)}{\Gamma(1/2)} \frac{2}{q(q+1)})F(1/2 + q - 1, 1 + q - 1; 3 + q - 1, z)$$

$$+\ z\left(\frac{\Gamma(1/2+q)}{\Gamma(1/2)}\frac{2}{(q+1)(q+2)}\right)F(1/2+q,1+q;3+q,z)\}$$

(26)

For $z \approx 1$, the $K_q$ is given by the following equation which has a singularity for $q \geq 2$:

$$K_q = x\frac{(2(q-2))!}{(q-2)!}\frac{1}{2^{2q-5}}\frac{1}{(1-z)^{(2q-3)/2}}$$

(27)

Hierarchical scaling relates the q'th factorial moment $K_q$ to the second factorial moment $K_2$. A similar feature also applies to normalized moments defined by $k_q = K_q / < N >^q$. An hierarchical scaling relation would read: $k_q = A_q k_2^{q-1}$. The $A_q$ depends only on $q$ in a hierarchical scaling relation. The $< N >= K_1$. The following relation is valid near $z \approx 1$ and reads

$$K_q = \frac{(2(q-2))!}{(q-2)!}\frac{2^{2(q-2)}}{3^{2(q-2)}}\frac{(K_2)^{2q-3}}{(K_1)^{2q-4}}$$

(28)

Simpler hierarchical scaling relation exist for models with $\tau < 2$. For comparison, a negative binomial distribution has $k_q = \Gamma(q)(k_2)^{q-2} = ((q-1)!)(k_2)^{q-2}$. Moreover, this simple scaling holds for all $z$ in the range $0 < z < 1$. For $\tau > 2$, the hierarchical structure relation does not hold for all $z$ but only when $z \approx 1$, and, even then, a simple relationship between the reduced moments $k_q$ and $k_2$ does not exist where $< N >= K_1$ does not appear in the relation.

### 3.4 The critical opalescence case $a = 2/3, b = 1, c = 3$ and $\tau = 2 + 1/3$, $\gamma = 2/3$.

The hypergeometric model $F(a,1;3,z)$ can be used to study phenomena where the grand canonical ensemble $Z_{gc}$ has terms which fall with $k$ as $1/k^{3-a}$ or which have exponent $\tau = 3 - a$ in the range $2 < \tau < 3$. We briefly mention another example from the phenomena of critical opalescence where droplet sizes around a critical point of a liquid/gas phase transition fall also with a power. At the critical point, a first order liquid/gas phase transition becomes second order and has a set of critical exponents associated with power law fall off of droplet sizes, divergences of thermodynamic quantities and scaling laws connected with the absence of a length scale. In the vicinity of a critical opalescence point a strong enhancement of density fluctuations occurs. For a heavy ion collision, a similar critical behavior has been suggested in ref[32] with a characteristic enhancement in particle fluctuations. Mean field descriptions of critical opalescence have $\tau$ =2 +1/3. The exponent $\tau$ and another critical exponent $\delta$ are connected as $\tau = 2 + 1/\delta$. The exponent $\delta$ describes the behavior of the pressure $P$ near the critical point: $P \to P_C$ as the density $\rho \to \rho_C$, with the subscript $C$ being the critical point values of these quantities. Specifically, $P - P_C \sim (\rho - \rho_C)^\delta$ with $\delta = 3$ in mean field theories. For $\tau = 2 + 1/3$ the hypergeometric model would be $F(2/3,1;3,z)$ and the associated $Z_{gc}$ is

$$Z_{gc} = \exp(\frac{3}{2} x \frac{3(1-z)^{4/3} - 3(1-z) + z}{z}) = \exp(\frac{3}{2} x \frac{(1-t)(3t^2 + 2t + 1)}{t^2 + t + 1})$$
(29)

The $t = (1-z)^{1/3}$. At the critical opalescence point $z \to 1, t \to 0$, the number of clans or Poisson clusters $N_C = (3/2)x$ and $< N >= (9/2)x$. The mean number of particles per clan or Poisson cluster is then $n_C =< N > / N_C = 3$. The void scaling function $\chi \to 1/ n_C = 1/3$ as $B < N >\to \infty$. The fluctuations become large as $z \to 1, t \to 0$ and the associated compressibility starts to diverge from eq(A.5). The $< N^2 > - < N >^2 \sim$ $2x/(1-z)^{2/3} = 2x/t^2$ and $B < N >\sim (4/9)/(1-z)^{2/3} = (4/9)/t^2$ as $z \to 1, t \to 0$. The exponent $\gamma = 2/3$ in the critical opalescence case. Away from the point $z \to 1, t \to 0$, the $t$ or $z$ dependence, with $t = (1-z)^{1/3}$, of various quantities are given by the following: The $B < N >= (2/3)((1-t^2)/t^2) \cdot ((2 + 6t + 3t^2 + t^3)/(3 + 2t - t^2 - t^3))$ which is considerably more complicated than the corresponding quantity of eq(23) for the Bose-Einstein model. The $< N >= x(3/2)((1-t)(3 + 2t + t^2)/(1 + t + t^2))$ while $\chi = (1 + 2t + 3t^2)/(3 + 2t + t^2)$. These equations don't lead to a simple analytic expression for $\chi$ as a function of $B < N >$ such as that of eq(24) for the Bose-Einstein model. However, a $\chi$ versus $B < N >$ behavior can be obtained numerically since each value of $t$ determines both the void function $\chi$ and scaling variable $B < N >$ uniquely.

### 3.5 Comparison of $a, b = 1, c = 2$ with $a, b = 1, c = 3$, and Gauss contiguous hypergeometric models

The hypergeometric model ${}_2F_1(a,1;3,z)$, which includes the Bose-Einstein model and the critical opalescence model, has the following features. For $a$ in the range $0 < a < 1$, the $N_C = (2/(2-a))x$ and $< N >= (2/((1-a)(2-a)))x$ are finite at $z = 1$. The $n_C =$ $< N > / N_C = 1/(1-a)$ and the void scaling function $\chi \to (1-a)$ as $B < N >\to \infty$ at $z = 1$. The $< N^2 > - < N >^2 \to 2x/(1-z)^a$ and $B < N >\to (1-a)(2-a)/(1-z)^a$ as $z \to 1$. The exponent $\gamma$ that describes the enhancement factor $\delta N^2 \sim 1/(1-z)^\gamma$ is thus $\gamma = a$. The exponent $\gamma$ also determines the behavior of $B < N >$ near $z = 1$. The large $B < N >$ behavior of $\chi$ thus contains information about $a, \gamma$ and the related critical exponent $\tau$. As $a \to 1$, $\tau \to 2$, $< N >\to \infty$. For $a \neq 1$, $\tau \neq 2$, the $< N >$ is finite. These behaviors can be contrasted with the behavior for a Poisson distribution where $B < N >= 0$ and $\chi = 1$ for any $B < N >$.

A comparison can be made with models based on the hypergeometric model ${}_2F_1(\bar{a},1;2,z) = F(\bar{a},1;2,z)$, where $a \equiv \bar{a}$ to avoid confusing with the symbol $a$ that has been used in $F(a,1;3,z)$. For the ${}_2F_1(\bar{a},1;2,z)$ case, the void scaling function is simply $\chi = ((1+B<N>)^{1-\bar{a}} - 1)/((1-\bar{a})B<N>)$. The $<N^2> - <N>^2 \to \bar{a}xz^2/(1-z)^{\bar{a}+1}$. The $B<N> \to \bar{a}z/(1-z)$ and $<N> \to xz/(1-z)^{\bar{a}}$ as $z \to 1$. Thus the exponent $\gamma$ defined by $\delta N^2 \sim 1/(1-z)^{\gamma}$ is $\gamma = 1+\bar{a}$. The $B<N>$ diverges with a different exponent than $\delta N^2$. For $\bar{a} \to 1$, the negative binomial limit, $\chi = \ln(1+B<N>)/(B<N>)$. Hierarchical structure relations are valid for all z and are simple relations that read:

$$k_q = (\Gamma(\bar{a}+q-1)/\Gamma(\bar{a}+1))(k_2)^{q-1} \tag{30}$$

Besides the negative binomial limit $\bar{a} = 1$, the case $\bar{a} = 1/2$, or Glauber model [30], is also of essential interest. The probability count distribution for $\bar{a} = 1/2$ arose from field emission from Lorentzian line shapes. The photon count distribution was then used as a possible hadronic particle multiplicity distribution in ref[33]. Thus, the negative binomial model and the Glauber model differ in the critical exponent $\tau = 2 - a$, but have the same $b, c$. The $Z_N$ in the probability function $P_N = Z_N / Z_{gs}$ can be evaluated in term of tabulated polynomials. The $Z_N$ in the Glauber model can be written in terms of the confluent hypergeometric function $U$ as $Z_N = (2x)^{2N} U(N, 2N, 4x) z^N / N!$ The $Z_{gc} = \exp(2x(1-\sqrt{1-z}))$. The $Z_N$ for the negative binomial case can also be written as $Z_N = (-1)^{N+1} U[-N+1, x, 0] t^N / N!$.

Hypergeometric function, such as $F(a,b;c,z)$ and $F(a,b;c+1,z)$, that differ by one unit in one of the variables $a,b,c$ are called Gauss contiguous. Thus, the hypergeometric $F(a,1;3,z)$ model is contiguous to the hypergeometric $F(a,1;2,z)$ model; the Bose-Einstein model based on $F(1/2,1;3,z)$ is contiguous to the Glauber model based on $F(1/2,1;2,z)$. An extension of $F(a,1;3,z)$ to its contiguous partner, $F(a,1;4,z)$ ,results in a change in the asymptotic behavior of associated $C_k$ from $\sim xz^k / k^{3-a}$ to $\sim xz^k / k^{4-a}$. These change in the index $c$ by one unit from $c = 2$ to $c = 3$ and to $c = 4$ shift the critical exponent $\tau$ by one unit from $\tau = 2 - a$ to $\tau = 3 - a$ and to a value of $\tau = 4 - a$, respectively. The $F(1/2,1;c,z)$ cases have $C_k \to (c-1)!\, xz^k / (\sqrt{\pi} k^{c-1/2})$ as $k \to \infty$.

The more general $F(a,b;c,z)$ has an associated $C_k \to [\Gamma(c)/(\Gamma(a)\Gamma(b))] \cdot xz^k / k^{1+c-a-b}$ and, consequently an exponent $\tau = 1 + c - a - b$. Thus, varying the indices $a,b,c$ allows for a study of the $\tau$ dependence of various quantities. For $\tau$ in the range: $1 < \tau \le 2$, $<N>, \delta N^2 \to \infty$; $2 < \tau \le 3$, $<N>$ is finite, ,$\delta N^2 \to \infty$; $3 < \tau$, $<N>, \delta N^2$ are both finite.

### 3.6 Summary of hypergeometric models

Results of the previous sections showed that power law behavior occurs at points where $z = 1$. In thermodynamic descriptions $z$ is identified with the fugacity or chemical potential $\mu$ with $z = \exp(\beta\mu)$. As also noted, $C_k$ can be considered combinants that generate discrete probability distributions $P_N$. In this case the analog of the grand canonical partition function is a generating function $Z_{gen}$. The combinants can be used to obtain information about the mean $N$, or $< N >$, associated with the $P_N$. They also contain information about all higher moments such as the fluctuation and skewness or $\delta N^2 = < N^2 > - < N >^2 = < (N - < N >)^2 >$ and $Sk = < (N - < N >)^3 >$ using $< (N - < N >)^m > = \Sigma\ (N - < N >)^m P_N = \Sigma k^m C_k$. The simple result connecting the central moment $< (N - < N >)^m >$ to the m'th moment of the $C_k$ distribution, $\Sigma k^m C_k$, is valid only for m=2, 3. The mean and fluctuation are also given in eq(A.3a) and eq(A.4). The second factorial moment of $C_k$ is $< N^2 > - < N >^2 - < N > = \Sigma\ k(k-1)C_k$. Higher central moments (m=4,5..) involve more complicated connections with moments of the combinants. The set of combinants $\vec{C} = \{C_1, C_2, C_{31}, ...\}$ generates the probability distribution $P_N$: $\exp \Sigma_{k=1}^{\infty} C_k (u^k - 1) = \Sigma_{N=0}^{\infty} P_N u^N$. The extra $-1$ in the exponent of this equation is the normalization factor necessary so that $\sum_{N=0}^{\infty} P_N = 1$. The exponential factor $\exp(\sum_{k=1}^{\infty} C_k u^k) \equiv Z_{gen}(\vec{C}, u) = \sum_{N=0}^{\infty} Q_N u^N$ generates a set of functions $Q_N$, $N = 0,1,2,...,\infty$, such that $P_N = Q_N / Z_{gen}(\vec{C}, u = 1)$.

Table1 summarizes specific cases we have considered. The $U(N, 2N, 4x)$ is a confluent hypergeometric function. The probability distributions of other models which don't have a special function representation can always be calculated using the recurrence relation of eq(A.11). As previously noted, the negative binomial model is a special case of a hypergeometric model with $a = 1, b = 1, c = 2$. The Glauber model belongs in the same class of $a, b = 1, c = 2$ as the negative binomial model and has the values $a = 1/2, b = 1, c = 2$. The Bose-Einstein model has $a = 1/2, b = 1, c = 3$, while the critical opalescence case has the same $b = 1, c = 3$, but with $a = 2/3$. The Poisson limit is also a special limiting case: $a \to 0, b = 1, c = 2$.

**Table1. Combinants for various models**

| Combinant $C_k$ | Probability Distribution $P_N$ | Remarks |
|---|---|---|
| $x\frac{[a]_{k-1}[1]_{k-1}}{[2]_{k-1}k!}z^k$ | | Hypergeometric: ${}_2F_1(a,1;2,z)$<br>$b=1, c=2$, $C_k \to xz^k/k^{2-a}$ |
| $C_1\delta_{k,1}$ | $C_1^N \exp(-C_1)/N!$ | Poisson with $<N>=C_1$ |
| $xz^k/k$ | $\frac{(N+x-1)!}{N!(x-1)!}\frac{(<N>/x)^N}{(1+<N>/x)^{N+x}}$ | Negative Binomial |
| $\frac{x}{2^{2(k-1)}}\frac{1}{k}\frac{(2(k-1))!}{((k-1)!)^2}z^k$ | $\frac{(2x)^{2N}U(N,2N,4x)z^N/N!}{\exp(1-2x\sqrt{1-z})}$ | Glauber Model<br>$C_k \to xz^k/k^{3/2}$ |
| $x\frac{[a]_{k-1}[1]_{k-1}}{[3]_{k-1}k!}z^k$ | | Hypergeometric: ${}_2F_1(a,1;3,z)$<br>$b=1, c=3$, $C_k \to xz^k/k^{3-a}$ |
| $x\frac{[1/2]_{k-1}[1]_{k-1}}{[3]_{k-1}k!}z^k$ | | Bose-Einstein Model<br>$a=1/2, b=1, c=3$, $C_k \to xz^k/k^{5/2}$ |
| $x\frac{[2/3]_{k-1}[1]_{k-1}}{[3]_{k-1}k!}z^k$ | | Critical Opalescence Model<br>$a=2/3, b=1, c=3$, $C_k \to xz^k/k^{7/3}$ |

---

Table 2 summaries various quantities that are considered. For the case $a, b=1, c=3$, the limiting behavior as $z \to 1$ is only given. The $\gamma$ is the exponent that describes the divergence of $\delta N^2 \sim 1/(1-z)^{\gamma}$ as $z \to 1$. The skewness $Sk =< (N- <N>)^3 >$ for $z \to 1$ are as follows. For $a, b=1, c=2$, the $Sk \to xa(1+a)/(1-z)^{2+a}$; for $a, b=1, c=3$, $Sk \to 2xa/(1-z)^{1+a}$.

---

**Table 2 . Behavior of various quantities**

| Model | $< N >$ | $\delta N^2$ | $\chi$ |
|---|---|---|---|
| $a,b=1, c=2$<br>$\tau = 2-a, \gamma = 1+a$ | $\frac{xz}{(1-z)^a}$ | $\frac{xz(1+(a-1)z)}{(1-z)^{a+1}}$ | $\frac{(1+B<N>)^{1-a}-1}{(1-a)B<N>}$ |
| $a,b=1, c=3$<br>$\tau = 3-a, \gamma = a$ | $\rightarrow \frac{2x}{(1-a)(2-a)}$ | $\rightarrow \frac{2x}{(1-z)^a}$ | $\rightarrow (1-a)$ |

## 4. Pion lasers, atoms in a laser trap, black body photons and pions, models from quantum optics.

### 4.1 Pion laser

In this section, we present some results on an exactly soluble model known in the literature as the pion laser. The reader is refered to ref [10] for details and derivations. Here we only give the final results for the quantities that are necessary for this paper. We also compare the results with atoms in a laser or harmonic oscillator trap and black body photons. The combinant for the pion laser model is

$$C_k = \frac{1}{k} \frac{(\frac{n_0}{\widehat{n}_c})^k}{(1-(\sqrt{\gamma_- / \gamma_+})^k)^3} \tag{31}$$

with $n_c = \gamma_+^{3/2}$, $\gamma_\pm = (1/2)(1+y\pm\sqrt{1+2y})$, and $y = R_e^2 \sigma_T^2$. The $\sigma_T^2 = \sigma^2 + 2mT$, $R_e^2 = R^2 + mT/(\sigma^2\sigma_T^2)$, and $\sigma^2$ = the width of the Gaussian wavepacket. The $n_0$ is the initial pion density which becomes enhanced by symmetrization effects. The $\widehat{n}_c$ is the critical density and in the notation used above, the $z = (n_0 / \widehat{n}_c)$. The critical condition is $z \rightarrow 1$, and $C_k$ falls as a pure power law and is not exponentially suppressed by $z^k = (n_0 / \widehat{n}_c)^k$. Furthermore, we mention that the one-dimensional version of the pion laser model has a closed form for $Z_A$. In one dimension the $n_c = \gamma_+^{1/2}$ and $(1-(\gamma_- / \gamma_+)^{k/2})^3 \rightarrow (1-(\gamma_- / \gamma_+)^{k/2})^1$. The partition function is:

$$Z_A = \frac{(\frac{n_0}{\hat{n}_c})^A}{\prod_{k=1}^{A}(1-q^k)} = \frac{(\frac{n_0}{\hat{n}_c})^A}{(1-q)(1-q^2)...(1-q^A)} \tag{32}$$

where $q = (\gamma_- / \gamma_+)^{1/2}$. The probability of having $A$ pions is $P_A = Z_A / \exp(N_C)$, where $N_C = \sum_{k=1}^{\infty} C_k = \sum_{k=1}^{\infty}(1/k)(n_0/\hat{n}_c)^k/(1-(q)^k)$. Two limits are interesting which are determined by the $\gamma_- / \gamma_+$ ratio. If $\gamma_- / \gamma_+ \to 0$, which happens when $y << 1$ (dilute, high $T$ gas), the distribution becomes a negative binomial independent of the number of dimensions. If $\gamma_- / \gamma_+ \to 1$, realized when $\sqrt{y} >> 1$ (dense, low $T$ gas) then for $\gamma_- / \gamma_+ = 1-\eta$, $(1-(\gamma_- / \gamma_+)^{k/2})^d \to (1-(1-\eta)^{k/2})^d \to (\eta k/2)^d$, where $d$ is the number of dimensions. Thus, the $C_k$ will fall with $k$ as $1/k^{1+d}$. Specifically, for $d = 1$, 2 and 3 dimensions, the $C_k$ will depend on $k$ as $1/k^2$, $1/k^3$, and $1/k^4$ respectively. The $d = 1$ case is the only one were the $Z_A$ can be expressed in closed form. The $d = 2,3$ cases can be obtained from the recurrence relation on $Z_A$.

## 4.2 Atoms in a laser trap

Atoms in a laser trap have a $C_k$ that can be obtained from [33]:

$$C_k = \frac{1}{k}\sum_{E_j}\exp(-k\beta E_j)$$

(33)

The $E_j$ are the single particle energies in the system and $\beta = 1/T$. For a harmonic oscillator trap: $E_j = (n_x + 1/2)\hbar\omega_x + (n_y + 1/2)\hbar\omega_y + (n_z + 1/2)\hbar\omega_z$. For equal $\omega's$, the resulting $C_k$ obtained by doing the series sums of eq(30) give

$$C_k = \frac{1}{k}\frac{((\sqrt{Q})^k)^3}{(1-Q^k)^3}$$

(34)

The $Q = \exp(-\beta\hbar\omega)$. For $d$ dimensions, $3 \to d$ for the 3 that appears in this last equation. The structure of eq(31) and eq(34) is somewhat similar. The result of eq(31) has an extra variable or control parameter $n_0 / n_c$.

## 4.3 Black body photons

As a final example, we consider black body photons in a volume $V$ and temperature $T$ which have a

$$C_k = g_s \frac{VT^3/\pi^2}{k^4}$$
(35)

The $g_s = 2$. The chemical potential $\mu = 0$ for photons and there is no Bose-Einstein condensation except at $T = 0$. Writing $C_k$ as $xz^k/k^\tau$, the $x = g_s VT^3/\pi^2$, $z = 1$. The $C_k$ fall as a pure power law with a $1/k^4$ dependence and thus $\tau = 4$. This $1/k^4$ behavior has the following consequences. The $<N> = \sum kC_k = g_s VT^3\zeta(3)/\pi^2$ and $<N^2> - <N>^2$ = $(\varsigma(2)/\zeta(3)) <N>$ are finite with the fluctuation simply proportional to $<N>$. From eq(A.5), the compressibility of a photon gas is finite or non-singular since the fluctuation is finite. The exponent $\gamma$ associated with a divergence in the fluctuation is $\gamma = 0$. However, the skewness $<(N - <N>)^3> = \sum k^3 C_k \propto \zeta(1) = \infty$ diverges. The $k$ dependence of $C_k$ for a system with dimension $d$ is $1/k^{1+d}$ and thus the exponent $\tau = 1 + d$. This $d$ dependence of $C_k$ can be compare with a non-relativistic gas which is $1/k^{1+d/2}$.

### 4.4 Thermal pions

In a zero mass limit, thermal black body pions will have features that parallel the photon result of the previous subsection. The photon combinant $C_k$ is simply extended to pions by replacing the spin degeneracy factor $g_s = 2$ for photons by $g_s = 1$ for each charged state or isospin state of the pion. When the mass of the pion is not negligible compared to $T$, the combinant is [33]:

$$C_k = \frac{VT^3}{2\pi^2}\left(\frac{m}{T}\right)^2 \frac{1}{k^2} K_2\left(k\frac{m}{T}\right) \rightarrow \frac{V}{(\lambda_T)^3}\frac{1}{k^{5/2}}\exp\left(-k\frac{m}{T}\right)$$
(36)

with $\lambda_T = h/(2\pi mT)^{1/2}$. The last part of this equation applies when $m >> T$, which can be obtained from the limiting behavior of the Bessel $K_2$ function. The zero mass limit reduces to the black body result of the previous section. Even when $m$ is not $<< T$, a $k$ will eventually be reached where $km/T >> 1$ and the $\rightarrow$ limit of the last equation applies for any $m \neq 0$. The $\exp(-km/T)$ exponentially truncates the distribution of $C_k$ and reduces the moments of it which give the mean, fluctuation, and skewness. Writing $C_k = xz^k/k^\tau$, this limit has $\tau = 5/2$, $x = V/(\lambda_T)^3$ and $z = \exp(-m/T)$.

### 4.5 Glauber quantum optics laser models.

In section 3.5 we mentioned that the special choice $\overline{a}$ =1/2, b=1,c=2, arises also in quantum optics where field emission of photons comes from states with Lorentzian line shapes. In this subsection we specify the meaning of the variables $x, z$ that appear in combinant $C_k \sim xz^k/k^{3/2}$ for this case. The variables $x, z$ are connected to properties of the field emission via the equations $x = T\Omega/2$ and $z = 2W\gamma/(2W\gamma + \gamma^2)$ in the notation of the first reference in ref[11]. The $\Omega = (\gamma^2 + 2W\gamma)^{1/2}$ and the $W$ is an integral over the Lorentzian line shape $\Gamma(\omega)$ which is $\Gamma(\omega) = b/[(\omega - \omega_0)^2 + \gamma^2]$. The $T$ is the time. The quantity $2x\sqrt{1-z}$ which appears in the model is just $\gamma T$. The reciprocal $\gamma^{-1} \equiv t_{coh}$ is the coherence time. The $< N >= xz/(1-z) = WT$ so that the mean count rate is $W =< N > /T$. The fluctuation is determined by a quantity $\mu = W/\gamma = t_{coh}W =< N > /\gamma T$ and is given by $< N^2 > - < N >^2 =< N > (1+\mu) =< N > (1+ < N > /\gamma T)$. The photon count distribution can be written as

$$P[N,T] = \frac{1}{N!}\sqrt{\frac{2\Omega T}{\pi}}\left[\frac{\gamma WT}{\Omega}\right]^N K_{N-1/2}(\Omega T)\exp(\gamma T)$$

(37)

## 5. Summary and conclusions.

In this paper we developed a model of particle multiplicity or count distributions based on a parallel with Bose-Einstein condensation that also has features related to critical phenomena, such as critical opalescence. This parallel led us to develop a generalized model which contained a parameter $a$, or equivalently, a critical exponent $\tau$. Various count distributions differ only by the choice of $a$ or $\tau$. The $a$ or $\tau$ were continuous parameters in the approach. We also introduced an associated critical exponent $\gamma$ which describes enhancements of fluctuations as a critical point in the theory is approached. Several commonly used models, such as the negative binomial model and a model in quantum optics originally due to Glauber and also used in particle physics, are special cases of our generalized approach. A detailed study of void scaling properties and features associated with hierarchical structure was carried out using the statistical framework developed in this paper. Such properties and features are useful not only in particle physics but also in other areas such as in cosmology where they are frequently used to analyze the large scale structure of the universe. Void scaling and hierarchical structure were first studied in the literature using a negative binomial distribution for count probabilities. For this distribution simple relations exist for the void function, called $\chi$, in terms of the void scaling variable which in this paper is labeled $B < N >$. The $B < N >$ is determined by the behavior of $< N^2 > - < N >^2 =< N > (1 + B < N >)$, and it is therefore related to the second cumulant $K_2$ through $B < N > = K_2 / < N >$. For the negative binomial distribution and for several other cases discussed, $\chi \to 0$ as $B < N > \to \infty$. As was shown in this paper, the property $\chi \to 0$ as $B < N > \to \infty$ is not a general property, but was found in our study to depend on a critical exponent $\tau$. For a class of distributions based on a hypergeometric function $F(a,1;2,z)$ which have $\tau \le 2$ and $\tau = 2 - a$, the void scaling function $\chi \to 0$ when $B < N > \to \infty$. The $\tau$ can be related to the Levy index $\alpha$ through the relation $\tau = 1 + \alpha$ or $a = 1 - \alpha$, where the Levy index $\alpha$ is restricted to the range $0 \le \alpha \le 1$. The negative binomial limit is $\alpha = 0, a = 1, \tau = 1$. Another important model, developed by Glauber for photon count distributions in quantum optics has $a = \alpha = 1/2, \tau = 3/2$. In addition, the negative binomial and more generally, for a class of models with index $\alpha \le 1$ or $1 \le \tau \le 2$, an hierarchical structure property is also present. This property relates the higher factorial moments of the combinants $C_k$ to powers of the second factorial moment.

A detailed study of a model based on a hypergeometric function $F(a,1;3,z)$ with an index $\tau > 2$ and $\tau = 3 - a$ was carried out to see what features still remain. Moreover, we were motivated to consider this extension of $\tau$ into the domain $\tau > 2$ by the fact that many phenomena in the physical sciences have $\tau > 2$. Of particular relevance to this paper was the fact that $\tau = 5/2$ for Bose-Einstein condensation of particles confined to a box in three dimensions. Our Bose-Einstein model was based on using $F(1/2,1;3,z)$ in the generating function of the count probability distribution and had $\tau = 5/2$ for the exponent in the power law fall off for the combinants. The $F(1/2,1;3,z)$ representation gave simple analytic results for the void distribution function in the variable $B < N >$ and also for the hierarchical structure functions. In this model the zeta functions that appear in the exact description of Bose-Einstein condensation are replaced with simple ratios of integers such as: $\varsigma(3/2) = 2.61 \to 8/3$ and $\varsigma(5/2) = 1.34 \to 4/3$. For this model we found that void scaling behavior is a property of the multiplicity distribution, but $\chi \to 1/2$ as $B < N > \to \infty$. This is in contrast to models with $\tau \le 2$ where $\chi \to 0$, as mentioned above. The asymptotic behavior of $\chi$ with increasing $B < N >$ is linked to the value of $\tau$ which determines either the exact behavior or the asymptotic behavior of the $C_k$'s. The number of Poisson clusters or clans $N_C$ is just the sum of the $C_k$ over k (zero moment), and the mean number of particles per Poisson cluster or clan $\equiv n_C$ is the ratio of the first to zero moment. The reciprocal of $n_C$ as $z \to 1$ is the limiting value of $\chi$ as the scaling variable $B < N > \to \infty$. For example, for the negative binomial $C_k = xz^k / k \sim 1/k$ for all $k$ and $z \to 1$ and $n_C \to \infty$, so that $\chi$ vanishes with increasing $B < N >$. For our Bose-Einstein model $n_C \to (8/3)/(4/3) = 2$ for $z \to 1$ and, thus, $\chi \to 1/2$ with increasing $B < N >$. The $\tau = 2$ is a transition value in $n_C$ from infinite values ($\tau \le 2$) to finite values ($\tau > 2$). Other cases of the asymptotic value of $\chi$ were also considered. A model based on $\tau = 2 + 1/3$ was also considered because of its relation to critical point behavior. This particular value of $\tau$ appears in a mean field theory of critical opalescence. The phenomena of critical opalescence arise from large density fluctuations that occur near the end point of a first order liquid/gas phase transition. At this end point a first order liquid/gas phase transition becomes a second order phase transition and the compressibility diverges. A similar critical point behavior for hadronic matter has been proposed [32].

The determination of critical exponents is important both from the theoretical and experimental standpoints. The exponent $\tau$ governs the properties of the void scaling function which can be used to obtain its value from experiment. The shape of the void scaling function $\chi$ as determined by its dependence on the scaling variable $B < N >=$ $K_2 / < N >$ can be used to obtain $\tau$. Moreover, the end point value of $\chi$ as $B < N > \to \infty$, $z \to 1$, is also useful in this determination for $\tau > 2$. The $\tau > 2$ in a hypergeometric $F(a,1;3,z)$ model, where $\tau = 3 - a$ and where $\chi \to 1 - a$ with $B < N > \to \infty$.

While void scaling was a feature of the several models studied in this paper, an hierarchical property was not found in general. It was only in the limit $z \to 1$, that hierarchical behavior was seen in the Bose-Einstein case where $\tau = 5/2$. Thus, we conclude that void scaling does not in general imply an hierarchical behavior. However, an hierarchical structure always leads to void scaling. Both hierarchical behavior and void scaling were present in the models with Levy index $0 < \alpha \le 1$ or $1 \le \tau < 2$ that we considered.

## Appendix A . Bose-Einstein condensation; Feynman's cycle class description.

### A.1 General remarks

The basic equations for Bose-Einstein (BE) condensation can be found in most textbooks in statistical physics[19,20]. A more elaborate description in terms of permutations, cycle classes and Feynman's density matrix approach to statistical mechanics can be found in ref[35]. In the next subsection we briefly summarize both approaches. In the process we will introduce some essential expressions and framework that will be used later. A d-dimensional volume will be considered so that $V_d = L^d$, with $L$ the length of one side of this volume. This will show how a critical exponent $\tau$ that appears in the theory can vary and in this case $\tau$ is related to the number of dimensions. In most phase transition, the exponent $\tau$ is unknown and has to be determined by experiment. Textbooks use the grand canonical ensemble, but methods based on the canonical and microcanonical ensemble have been developed for BE condensation [33] both in a box as a confining volume and in the laser or harmonic oscillator trap. Solutions in the harmonic oscillator trap [33] have some formal similarities with solutions of the pion laser model [10] which will also be briefly mentioned.

### A.2 Standard description

The BE occupation factor of an energy level $\varepsilon$ is $1/(\exp(\beta(\varepsilon-\mu))-1)$. The $\mu$ is the chemical potential and $\beta = 1/T$. The $\mu \le \varepsilon_{gs} = 0$ where $\varepsilon_{gs}$ is the ground state single particle energy which is taken as zero. The density of states for particles in a d-dimensional volume is $L^d d^d p / h^d$. The mean number of particles follows from integrating the occupation factor over the density of states with $\varepsilon = p^2/(2m)$ in the non-relativistic limit. This is achieved by using $1/(y-1) = 1/(y(1-y^{-1}) = \sum_{k=1}^{\infty} y^{-k}$ or $1/(\exp(\beta(\varepsilon-\mu))-1) = \sum_{k=1}^{\infty} \exp(-k\beta(\varepsilon-\mu))$ and doing the Gaussian integrals over $p$. The Feynman approach gives a physical meaning to this expansion and identifies the index $k$ in this expansion with the length of a cycle as will be discussed below. The resulting equation for the mean number of particles is

$$< N >= (L^d / \lambda^d) \sum_{k=1}^{\infty} z^k / k^{d/2}$$

(A.1)

The $\lambda = h/(2\pi mT)^{1/2}$ is the thermal wavelength. The fugacity $z = \exp(\beta\mu)$. The fugacity $z \to 1$ when $\mu \to 0$. At $z = 1$ the sum that appears in eq(1) is $\varsigma(d/2)$. The $\varsigma(d/2)$ converges for $d > 2$ and diverges $d \le 2$. For 3 dimensions $\varsigma(d/2) = \varsigma(3/2) = 2.61$. In 1 and 2 dimensions $\varsigma(d/2) = \infty$. The grand canonical partition function is

$$Z_{gc} = \exp(\,(L^d/\lambda^d)\sum_{k=1}^{\infty} z^k/k^{(1+d/2)}) = \exp(PV_d/T)$$

(A.2)

Here $P$ is the pressure and $V_d = L^d$. Setting $C_k = xz^k/k^{(1+d/2)}$, where $x = L^d/\lambda^d$, then

$$< N >= \sum_{k=1}^{\infty} kC_k = x\sum_{k=1}^{\infty} z^k/k^{d/2}$$

(A.3a)

$$Z_{gc} = \exp(\sum_{k=1}^{\infty} C_k) = \exp(x\sum_{k=1}^{\infty} z^k/k^{1+d/2})$$

(A.3b)

The $< N >= z(\partial Z_{gc}/\partial z)/Z_{gc}$. Similarly, the fluctuation $< N^2 > - < N >^2$ is given by

$$< N^2 > - < N >^2 = \sum_{k=1}^{\infty} k^2 C_k = x\sum_{k=1}^{\infty} z^k/k^{d/2-1}$$

(A.4)

The particle number fluctuations can be related to the isothermal compressibility coefficient $\kappa_T = -(1/V)(\partial V/\partial P)_T$. Namely,

$$< N^2 > - < N >^2 = < N > T\kappa_T/v$$

(A.5)

Here, $v = V/< N >$. In 3 dimensions $< N^2 > - < N >^2 \to \infty$ at $z$=1 since $C_k \sim 1/k^{5/2}$. This divergence in $\delta N^2 = < N^2 > - < N >^2$ is associated with the singularity of the compressibility $\kappa_T$ at $z$=1. Near $z$=1, small changes in the pressure lead to large changes in the volume. While the specific heat at constant volume has a cusp [19-21], the specific heat at constant pressure becomes singular since the two specific heats are related through a compressibility factor or thermal expansion coefficient which is also singular. The fluctuation in the number of bosons in an energy level $\varepsilon$ is given by the Planck result $< n_\varepsilon{}^2 > - < n_\varepsilon >^2 = < n_\varepsilon > (1 + < n_e >)$ where $< n_\varepsilon > = 1/(\exp(\beta(\varepsilon - \mu)) - 1)$. In one dimension the power law fall off that appears in $< N >$ is $kC_k \sim k/k^{3/2} = 1/k^{1/2}$ at $z$=1. Thus, in one dimension, $< N >$ also diverges. In particle multiplicity distributions properties associated with fluctuations are of essential interest. The connection of eq(A.5) between $\kappa_T$ and $\delta N^2$ offers an important link between thermal properties and particle multiplicity distributions.

The $C_k$ are called combinants [36] in approaches having to do with multiplicity distributions. The $Z_{gc}$ will be related to the generating function for the probability distributions associated with the multiplicity. The results of eq(A.3a,3b) and eq(A.4) also appear in the theory of multiplicity distributions. Different multiplicity distributions are generated with different exponents for the power law prefactor part of the $C_k$. The $x = L^d / \lambda^d$, which is an extensive quantity in statistical descriptions will no longer be extensive quantity in general count probability distributions. It will be a parameter that can control the size of the fluctuation. For a negative binomial distribution $C_k = xz^k / k$, where $x$ is the negative binomial parameter and $< N^2 > - < N >^2 = < N > (1 + < N > / x)$.

The conditions for BE condensation are obtained from eq(A.3a) and the feature that $\mu \to 0, z \to 1$ at the condensation point. At volume and/or temperature conditions above the critical point, the $z < 1, \mu < 0$. As the critical condition is approached by varying the $V_d = L^d$ and/or the temperature $T$ (present in $\lambda$), the $z$ will approach the value 1. The largest value of $\sum z^k / k^{d/2}$ is $\varsigma(d/2)$ which is realized at $z$ =1. For $d = 3$, $\varsigma(3/2) = 1.34$. For $d = 2$, $\varsigma(2/2) = \infty$. Next, let us consider the right hand side of the equation: $< N > /(L^d / \lambda^d) = \sum_{k=1}^{\infty} z^k / k^{d/2}$ which is just eq(A.1) rearranged. At fixed volume $V_d$, lowering $T$ or increasing $\lambda$, increases the left hand side of this equation. For $d = 1,2,$ this constraint equation can always be satisfied since the right hand side can increase indefinitely. Condensation will still occur but now at $T$ =0. However, for $d = 3$ the right hand side can only become $\varsigma(3/2) = 1.34$. The lowest $T = T_c$ satisfies the equation $(< N > / V)(2\pi m T_c)^{3/2} / h^3 = \varsigma(3/2)$. Similarly, for fixed $T$, a smallest $V = V_c$ is reached when $(< N > / V_c)\lambda^3 = \varsigma(3/2)$. Below this $T_c$ or less then this $V_c$, condensation occurs. The constraint of eq(A.1) can no longer be maintained and the ground state occupancy must be taken out of the sum to give the well known result: $< N_{gs} > = < N > (1 - (T / T_c)^{3/2})$.

One final important connection relates the canonical ensemble $Z_A$ and grand canonical ensemble through the relation:

$$Z_{gc} = \sum_{A=0}^{\infty} Z_A \exp(\beta \mu A)$$

(A.6)

The $Z_0 = 1$ and is the system with no particles which will be related to the void probability when we discuss void scaling in multiplicity distributions.

A.3 Feynman density matrix - cycle index approach

We now briefly mention the application of Feynman's density matrix approach in statistical mechanics to BE condensation phenomena. The reader is referred to ref [35,6] for further details. The canonical partition function $Z_A$ for A particles is obtained from the density matrix and involves a symmetrization operator over A particles. The symmetrization operation brings in A! permutations which can be broken into a cycle class representation. For example, the permutation:

$$\begin{bmatrix} \frac{12345678910}{21357648109} \end{bmatrix}$$

(A.7)

or $1 \to 2 \to 1, 3 \to 3, 4 \to 5 \to 7 \to 4, 6 \to 6, 8 \to 8, 9 \to 10 \to 9$, can be represented by specifying the number $n_k$ of cycles of length k . The above permutation has $n_1 = 3, n_2 = 2, n_3 = 1$ and all other $n_k = 0$ . Each permutation can be represented by a vector $\vec{n} = (n_1, n_2, n_3, ..., n_A)$ subject to $A = \sum kn_k$ . For the $A!$ permutations, $M_2(\ n_1, ..., n_A)$ $= A!/\prod k^{n_k} n_k!$ belong to the cycle class $(n_1, n_2, n_3, ..., n_A)$. The cycles form loops and represent the linkage of various elements to each other such as $4 \to 5 \to 7 \to 4$ . The linkages or permutations have the following properties. The loops are closed, starting and ending on one of the members of the cycle. The starting point of a cycle can be any one of the k members and this feature accounts for the $1/k$ term in $M_2$ for each $n_k$ . Specifically, the following cycles are the same: $4 \to 5 \to 7 \to 4$ , $5 \to 7 \to 4 \to 5$, and $7 \to 4 \to 5 \to 7$ . Rearrangements of members within a cycle leads to a new permutation. For example, the cycles $4 \to 5 \to 7 \to 4$ and $4 \to 7 \to 5 \to 4$ are different permutations. Every partition $\vec{n}$ is then given the following weight

$$W_A(\vec{n}, \vec{c}) = \coprod_{k=1}^{\infty} M_2(\vec{n})(c_k)^{n_k} / n_k!$$

(A.8)

The lower case $c_k = xz^k / k^{3/2}$ where $x = V / \lambda^3$ . The overall $A!$ in $M_2$ can be dropped and the $1/k$ term in $M_2$ can be incorporated into $c_k = xz^k / k^{3/2} \to xz^k / k^{5/2} \equiv C_k$ , defining an upper case $C_k$ . The resulting $W_A(\vec{n}, \vec{C})$ is eq(A.8) with $M_2$ now absent and has $c_k$ replaced with $C_k$ . The canonical ensemble partition function is obtain as a sum of $W_A(\vec{n}, \vec{C})$ over all $\vec{n}$ :

$$Z_A(\vec{C}) = \sum_{\vec{n}} W_A(\vec{n}, \vec{C})$$

(A.9)

The canonical ensemble has a fixed number of particles with the constraint $A = \sum kn_k$ .

A grand canonical partition function can be generated from the canonical ensemble of eq(A.9) using the fugacity $z$ as a weight factor as in eq(A.6). The mean of cycles of length k, or $< n_k >$, in the canonical ensemble is given

$$< n_k >= \sum_{\vec{n}} \prod_k n_k (C_k)^{n_k} / n_k! / Z_A(\vec{C}) = C_k \sum_{\vec{n}} \coprod_k (C_k)^{n_k - 1} /(n_k - 1)! / Z_A(\vec{C}) =$$

$$C_k Z_{A-k}(\vec{C}) / Z_A(\vec{C})$$

(A.10)

The last equality in eq(A.10) arises from the observation that the sum over all partitions of $\vec{n}$ is now over a system with $n_k \to (n_k - 1)$. The result of this sum on the reduced system of one less $n_k$ is the canonical partition function for a system of size $A - k$. The constraint $A = \sum k n_k = \sum k < n_k >$ and the result of eq(A.10) leads to the recurrence relation of ref[37]:

$$Z_A(\vec{C}) = (1/A) \sum_{k=1}^{A} k C_k Z_{A-k}(\vec{C})$$

(A.11)

The recurrence relation is very useful in calculating the canonical partition function. In some special cases for $C_k$, the resulting $Z_A(\vec{C})$ becomes a simple function. These cases will be given. Thermodynamic quantities are contained in $Z_A$ through the connection:

$$Z_A(\vec{C}) = \exp(-\beta F_A(V,T))$$

(A.12)

The $F_A(V,T)$ is the Helmholtz free energy and $dF_A(V,T) = SdT - PdV + \mu dA$, where the entropy $S$ and other thermodynamic quantities that appear in this equation can be calculated from partial derivatives such as $S = (\partial F_A(V,T) / \partial T)_{V,A}$. The discrete version of $dF_A(V,T)$ gives $\mu = F_{A+1}(V,T) - F_A(V,T)$ and $Z_{A-1} / Z_A = \exp(-\beta(F_{A-1} - F_A)) = \exp(\beta\mu)$. The chain $(Z_{A-1} / Z_A) \cdot (Z_{A-2} / Z_{A-1}) \cdot ... \cdot (Z_{A-k} / Z_{A-k+1}) = Z_{A-k} / Z_A = \exp(\beta\mu k)$. The $< n_k >$ in the grand canonical ensemble is $< n_k >= C_k \exp(\beta\mu k) = C_k z^k$. The recurrence relation of eq(A.11) is useful for studying how various quantities evolve as more and more particles are introduced or as $A$ increases. The cycle lengths give information about the degree of linkage or correlations. The presence of long cycles reflects strong correlations between many particles or members of the ensemble. The high temperature limit is all unit cycles and corresponds to a situation of no statistical correlations. In this limit, a Maxwell-Boltzmann distribution applies and Poisson statistics is realized. By comparison, at condensation $\mu \to 0, z \to 1$ and $< n_k > \to C_k$. Thus, the $< n_k >$ fall as a pure power law with no exponential damping from $z^k = \exp(\beta\mu k)$. Above the condensation point, $\mu < 0$.

### A.4 Summary of relevant results

In the process of summarizing the two approaches to BE condensation we have introduced various important quantities such as the combinant $C_k$, the results of eq(A.3a,A.3b), The importance of a power law prefactor exponent $\tau$ in $C_k \sim 1/k^\tau$ was stressed. For BE condensation in $d$ dimensions, the $\tau = 1 + d/2$. Thus $\tau$ depends on the dimensions of the system. In general, $\tau$ is a critical exponent [19] that manifests itself at a 2nd order phase transition and is not restricted to integer or half integer values. We will therefore treat it as a variable. Also discussed were the canonical partition function $Z_A$ and its recurrence relation, the grand canonical partition function $Z_{gc}$ as a generating function for $Z_A$, and the underlying weight $W_A(\vec{n},\vec{C})$ over all partition vectors $\vec{n}$. These results are not specific to BE condensation. The Feynman cycle class picture leads to an understanding of the correlations or linkages in terms of the number of cycles and their lengths. The cycles can be pictured as topological loops or closed paths that end on themselves. The $1/k^{d/2}$ factor in $C_k$ represents a Brownian motion like term which reflects a return to its initial point in a network chain (closed loop or cycle in $d$ dimensions). The presence of very long chains or cycles that are not exponentially cut off, but which decay as a pure power law leads to singularities in the behavior of the system. For example, thermodynamic quantities such as the compressibility and the specific heat at constant pressure in BE condensation become infinite. A large enhancement in the compressibility is associated with a correspondingly large enhancement in density fluctuations from eq(A.5). In particle multiplicity distributions, the $< N >$ and variance $\delta N^2$ can become very large depending on the exponent $\tau$. Scaling laws associated with these singularities will be discussed. The value of the exponent $\tau = 1 + d/2$ and how it relates to BE condensation parameters such as the critical $T_C$ and/or critical $V_d$ was also mentioned. The connection to Levy distributions is contained in the fact that moments of the distribution can become infinite. For a Levy index $\alpha$ (which is between 0 and 1), the associated probability distribution has no finite moment except for the zero moment of the distribution. Stable Levy distributions contain a factor similar to $z^k = \exp(-yk)$ so that sums over k are finite. The BE critical exponent $\tau$ and the Levy index $\alpha$ are related as $\tau = 2 - \alpha$. For $\alpha$ in the range $0 < \alpha \le 1, \tau$ is in the range $1 \le \tau < 2$. In analogy, BE properties are stabilized by the fugacity $z = \exp(\beta\mu)$, $\mu < 0$. The case of d=1 dimensions has $C_k = xz^k / k^{3/2} \to x / k^{3/2}$ at the condensation point $z = 1$ and is similar to a distribution with Levy index $\alpha = 1/2$. In 1 dimension ,$< N >= \Sigma k C_k \sim \Sigma 1/k^{1/2} = \varsigma(1/2) = \infty$ at $z = 1$. Similarly, for d=2, $< N >= \Sigma k C_k \sim \Sigma 1/k^1 = \varsigma(1) = \infty$ at $z = 1$. An important consequence of these infinities is that the condensation occurs only at $T_C = 0$. The Bose-Einstein condensation, for d=3, has finite zero and first moment for the sums over the cycle class distribution $C_k \to x/k^{5/2}$ at condensation. The mean number of cycles of length k has the feature $< n_k > \to x/k^{5/2}$ at $z = 1$, and falls as a pure power law at condensation. The divergence in the second moment of the $C_k$ distribution, $\Sigma k^2 C_k \sim \Sigma 1/k^{1/2} = \varsigma(1/2) = \infty$ shows up as a divergence of the incompressibility and specific heat at constant pressure $P$ in this case. Since the value of $\tau$ plays such an important role in BE condensation, one of the main issues in this paper is to explore its role in particle multiplicity distributions. As we shall see, the value $\tau = 2$ is an important transition value for void scaling and hierarchical structure behavior.

As a final note in this section and as an example of a truncated Levy distribution consider card shuffles or random permutations. The weight given to any cycle class vector $\vec{n}$ describing a permutation is the quantity $M_2(\vec{n}) = M_2(n_1, n_2, ..., n_A)$ given in sect A.2.3 in the description of Feynman's discussion of BE condensation. The $C_k = 1/k$, and the mean number of cycles is $< n_k >= \Sigma n_k M_2(\vec{n}) / \Sigma M_2(\vec{n}) = 1/k$, $k = 1,2,3,....,A$ with $< n_k >= 0$, $k > A$. The $\Sigma$ 's are over $\vec{n}$ and $\Sigma M_2(\vec{n}) = A!$. The distribution of cycles of length $k$ is a pure power law truncated at the size of the system. Moreover, for this particular case the distribution of $< n_k >$ remains the same as $A$ increases, with the only effect of adding longer and longer cycles with increasing $A$.

The statistical models of particle production are simple generalizations of results in the last paragraph. Namely, the weight $M_2(\vec{n})$ is replaced by $W_A(\vec{n}, \vec{C})$ and the $1/k$ factor in $M_2(\vec{n})$ now becomes $C_k \sim xz^k / k^\tau$ in $W_A(\vec{n}, \vec{C})$. Count probability distributions $P_N$ will follow from $P_N = Z_N / \Sigma_{n=0}^{\infty} Z_N$.

## Acknowledgments

A.Z.Mekjian would like to thank the Fulbright Foundation for a Fulbright Scholars Grant to visit Hungary and KFKI were parts of this research were done. This research was also done in part under a DOE grant-DOE-FG02-96ER-40987

## References

[1] L.P.Pitaevskii and S.Stringari, Bose-Einstein Condensation (Oxford University Press, 2003)

[2] C.J.Pethick and H.Smith, Bose-Einstein Condensation in Dilute Gases (Cambridge University Press, 2002)

[3] R.Hanbury-Brown, R.Q. Twiss, Nature 117, 27 (1956); The Intensity Interferometer (Taylor&Francis, N.Y., 1974)

[4] R.M.Weiner, Phys. Rep. 327, 249 (2000); Introduction to Bose-Einstein Correlations and Subatomic Interferometry, (John Wilet&Sons, Ltd., 2000)

[5] T.Csörgő, Acta Phys. Hung., New Series, Heavy Ion Phys. 15, 1-80 (2002)

[6] K.Chase and A.Z.Mekjian, Phys. Rev. C49, 014907 (1994)

[7] G.Bianconi and A.L.Barabasi, Phys. Rev. Lett. 86, 5632 (2001)

[8] R.Albert and A.l.Barabasi, Rev. of Mod. Phys., 74, 47 (2002)

[9] S.Pratt, Phys. Lett. B301, 159 (1993)

[10] T.Csörgő and J.Zimányi, Phys. Rev. Lett. 80, 916 (1998), Heavy Ion Phys. 9, 24 (1999)

[11] J.R.Klauder and E.C.G.Sudarshan, Fundamentals of Quantum Optics, (WA Benjamin Inc., N.Y., 1968)
M.O.O.Scully, M.S.Zubairy, Quantum Optics, (Cambridge University Press, Cambridge, 1997)
[12] P.Carruthers and C.C.Shih, Int. J. Mod. Phys. A 2, 1447 (1987)
[13] S.Hegyi, Phys. Lett. B309, 443 (1993); B414, 210 (1997); B411, 321 (1997); B466, 380 (1999): B467, 126 (1999)
[14] A.Z.Mekjian, Phys. Rev. Lett. 86, 220 (2001); Phys. Rev. C65, 014907 (2002)
S.J.Lee and A.Z. Mekjian, Nucl. Phys. A730, 514 (2004)
[15] Z.Koba, H.B.Nielsen and P.Olesen, Nucl.Phys.B40, 317 (1972)
[16] S.D.White, MNRAS 186, 145 (1979)
[17] S.Lupia, A. Giovannini and R.Ugoggioni, Z.Phys. C59, 427 (1993)
[18] W.Feller, An Introduction to Probability Theory (John Wiley, N.Y., 1971)
B.B.Mandelbrot, The Fractal Geometry of Nature, (W.H.Freeman, N.Y., 1982)
[19] K. Huang, Statistical Mechanics, 2nd ed. (John Wiley and Sons, N.Y., London, 1984)
[20] C.Kittel and H.Kroemer, Thermal Physics, 2nd ed. (Freeman, N.Y., 1980)
[21] T.Csörgő, S.Hegyi and W.Zajc, Eur. Phys. Journal C36, 67 (2004)
[22] D.J.Croton etal, Mon. Not. R. Astron. Soc. 352, 828 (2004)
[23] R.Balian and R.Schaeffer, A&A 220, 1 (1989)
[24] J.N.Fry, ApJ. 306, 358 (1988)
[25] S.N.Dorogovtsev and J.F.F.Mendes, Adv. Phys. 51, 1079 (2002); Evolution of Networks: From Biological Networks to the Internet and WWW, Oxford Univ. Press (2003)
[26] F.Gelis and R. Venugopalan, hep-ph/0605246, hep-ph/0601209 , hep-ph/0608117.
[27]I.M.Dremin and J.W.Gary, Phys. Rep 349, 301 (2001)
[28] E.A.De Wolf, I.M.Dremin and W.Kittel, Phys. Rept 270. 1 (1996)
[29] I.M.Dremin, hep-ph/0607183
[30] R.Glauber, Phys. Rev. Lett. 10, 84 (1963)
[31] P.Carruthers and I.Sarcevic, Phys. Rev. Lett. 63, 1562 (1989)
[32] M.Stephanov, K.Rajagopal and E.Shuryak, Phys. Rev. Lett. 81, 4816 (1998)
[33] K.C.Chase, A,Z.Mekjian and L.Zamick, Eur. Phys. Journal B8, 281 (1999)
[34] A.Z.Mekjian, B.R.Schlei and D.Stottman, Phys. Rev. C58 3627 (1998)
[35] R.P.Feynman, Statistical Physics, a set of lectures, (Addison Wesley, Boston, Mass.,1972)
[36] M.Gyulassy, S.K.Kauffman and L.Wilson, Phys. Rev. Lett. 40, 298 (1978)
[37] K.Chase and A.Z.Mekjian, Phys. Rev. C52, R2339 (1995)